\documentclass[twocolumn,showpacs,amsmath,amssymb,prl]{revtex4}
\usepackage{graphicx}
\usepackage{dcolumn}
\usepackage{bm}

\begin{document}

\title{Energetic Electrons and Nuclear Transmutations in Exploding Wires}

\author{A. Widom}
\affiliation{Physics Department, Northeastern University, Boston MA 02115, U.S.A}
\author{Y.N. Srivastava}
\affiliation{Dipartimento di Fisica \& INFN, Universita`degli Studi di Perugia, 
06123 Perugia, Italy}
\author{L. Larsen}
\affiliation{Lattice Energy LLC, 175 North Harbor Drive, Chicago IL 60601, U.S.A.}

\begin{abstract}
Nuclear transmutations and fast neutrons have been observed to emerge from large electrical 
current pulses passing through wire filaments which are induced to explode. The nuclear 
reactions may be explained as inverse beta transitions of energetic electrons absorbed 
either directly by single protons in Hydrogen or by protons embedded in other more massive 
nuclei. The critical energy transformations to the electrons from the electromagnetic field 
and from the electrons to the nuclei are best understood in terms of coherent collective 
motions of the many flowing electrons within a wire filament. 
Energy transformation mechanisms have thus been found which settle a theoretical paradox 
in low energy nuclear reactions which has remained unresolved for over eight decades. 
It is presently clear that nuclear transmutations can occur under a much wider range of 
physical conditions than was heretofore thought possible. 
\end{abstract}


\maketitle

Over eighty years ago, Wendt and Irion\cite{WendtIrion:1922} reported 
nuclear reactions in exploding wires. The transmuted nuclear products 
emerged after a large current pulse was passed through a Tungsten wire 
filament which exploded. Sir Ernest Rutherford\cite{Rutherford:1922} 
expressed doubts as to whether the electrons flowing through the wire 
could carry enough energy to induce nuclear reactions. The exploding 
wire current pulse had been produced by a capacitor discharge with an 
initial voltage of only thirty kilovolts. On the other hand, Rutherford 
had employed a high energy but dilute beam of 100 KeV electrons fired 
into a Tungsten target. Rutherford did not observe any nuclear reactions.
Wendt\cite{Wendt:1922} replied to the Rutherford objections, asserting 
that the peak power in the exploding wire current pulse was much larger 
than the relatively small power input to Rutherford's electron beam. 
Most importantly, a large energy transfer from the many electrons in 
the wire to the nuclei could occur collectively which would allow for 
the nuclear transmutation energy. 

This very old but important debate between Wendt and Rutherford has 
presently been experimentally settled in favour of Wendt. The more 
recent\cite{Stephanakis:1972,Young:1977,Bakshaev:2001,Bakshaev:2006,Velikovich:2007,Coverdale:2007} 
exploding wire experiments have, beyond any doubt, detected fast 
emerging neutrons capable of inducing nuclear transmutations. 
These observed fast neutrons have often been attributed to the products 
of deuteron fusion but we find that hypothesis unlikely to be true.
Firstly, fast neutrons have been seen in exploding wires 
even though there were no deuterons initially present \cite{Stephanakis:1972,Young:1977}. 
Secondly, the gamma emission signature of deuterium fusion has not been observed. 
It is much more likely that the fast neutrons are products of inverse 
beta transitions  of very energetic electrons being absorbed by protons 
and producing fast neutrons and neutrinos. The protons may be Hydrogen 
atomic nuclei or the protons may be embedded within more massive nuclei. 
The theoretical side of the difference of opinion between Wendt and 
Rutherford concerning how large amounts of energy can be transferred to 
and from the electrons in the wire has remained unresolved. 
The purpose of this work is to explain how this collective energy transfer 
may occur. 

The reader has perhaps unwittingly observed exploding Tungsten wire filaments 
in household light bulbs. Normally, the hot wire glows in the yellow optical 
frequency, slowly evaporating metal atoms from the filament. Over time, the 
wire filament thins in some pinch regions, strongly increasing the Maxwell 
magnetic pressure. Then with a "pop", the filament explodes, shifting the 
final bright radiation pulse frequency upward into the blue. The filament is 
broken at the pinch points. The reader may have shaken the bulb and heard the 
broken metallic pieces of filament rattle from within the bulb or noted the 
explosion dust on the inside surface of the bulb before replacing the old light 
bulb with a new one. Now let us consider this process only at much higher currents, 
temperatures, pressures and energies\cite{Sarkisov:2005}. 

The scale of wire currents required to induce nuclear reactions may be 
found by expressing the rest energy of the electron   
\begin{math} mc^2 \end{math}in units of a current 
\begin{math} I_0  \end{math}; i.e. by employing the vacuum impedance  
\begin{math} R_{vac}  \end{math} one finds 
\begin{equation}
\frac{mc^2}{|e|}=\frac{R_{vac} I_0}{4\pi }
\ \ \Rightarrow \ \ 
I_0\approx 1.704509\times 10^4\ \ {\rm Ampere}.
\label{MassCurrent}
\end{equation}
If a strong current pulse, large on the scale of \begin{math} I_0  \end{math}, 
passes through a thin wire filament, then the magnetic field exerts a very 
large Maxwell pressure on surface area elements, compressing, twisting and 
pushing into the wire. If the magnetic Maxwell pressure grows 
beyond the tensile strength of the 
wire material at the hot filament temperature, then the wire begins to melt 
and disintegrate. If the heating rate is sufficiently fast, then the hot 
wire may emit thermal radiation at a very high noise temperature. The 
thermal radiation for exploding Tungsten filaments exhibits X-ray 
frequencies indicating very high electron kinetic energies within the 
filament. Due to the electron kinetic pressure, the wire diameter starts 
to increase yielding a filament dense gas phase but still with some liquid 
droplets. The final explosive product consists of a hot plasma colloid 
containing some small dust particles of the original wire material. These 
products cool off into a gas and some smoke as is usual for explosions.

In order to understand how the electrons are accelerated into high energy 
regimes, recall that a very rapidly changing current induces a Faraday law 
voltage across an inductive circuit element. The Faraday law voltage per 
unit length \begin{math} E \end{math} is determined by the self inductance 
per unit length \begin{math}  \eta \mu_0 /4\pi  \end{math}, 
\begin{equation}
E=\eta \left(\frac{\mu_0}{4\pi }\right)\frac{dI}{dt}
=\frac{\eta }{c}\left(\frac{R_{vac}}{4\pi }\right)\frac{dI}{dt}\ .
\label{Inductance}
\end{equation} 
The dimensionless geometrical factor \begin{math} \eta \end{math} for the 
inductance per unit length varies very slowly with the wire cross sectional 
area \begin{math} A_0 \end{math} and the area   
\begin{math} A \end{math} along the circuit element collecting the magnetic 
flux, i.e. \begin{math} \eta =\ln (A/A_0)>1 \end{math}. The electric field 
\begin{math} E \end{math} will change the momentum \begin{math} p \end{math} 
and thereby the energy \begin{math} W \end{math} of a negatively charged 
electron \begin{math} e=-|e| \end{math}. The equation of motion is 
\begin{math} \dot{p}=eE  \end{math}. In virtue of Eqs.(\ref{MassCurrent}) 
and (\ref{Inductance}), one finds 
the central result of our work, i.e. the power 
\begin{math} e{\bf v\cdot E} \end{math} delivered to a moving electron 
by a changing current obeys
\begin{equation}
\frac{dW}{dt}=evE=-\eta mc^2\left(\frac{1}{I_0}\frac{dI}{dt}\right)
\frac{v}{c}\ .
\label{power}
\end{equation}
A change in the collective current \begin{math} dI \end{math} yields a changing 
single electron momentum and thereby a change in the single electron energy 
\begin{math} dW \end{math} wherein \begin{math} v \end{math} is the velocity 
of that electron. The single electron energy can thereby reach values far 
above the electron rest energy for a pulse peak current large on the scale 
of \begin{math} I_0 \end{math}.

The following comments are worthy of note: (i) The electromagnetic field 
configuration when the current pulse passes through the wire is a magnetic 
field tangent to the wire surface and normal to the wire axis and an electric 
field parallel to the cylinder. This is the low circuit frequency limit of the 
surface plasma polariton mode previously employed in the 
explanation\cite{Widom:2006} of inverse beta transitions in chemical cells. 
However, the natural surface patches whereon the long wavelength neutrons 
would form are in the case of thin wire filaments destroyed by the explosion.
(ii) Radiation losses have not been included in the above 
discussion. These losses are not large because of the collective nature of the current. 
A single charged accelerating particle emits copious radiation whereas many 
electrons contributing to a smooth current in a wire will hardly radiate at 
all. However, some resistive wire heating energy will be removed from the 
wire filament as hot emitted thermal radiation. (iii) The Maxwell electromagnetic 
energy and pressure are largely due to the Ampere's law mutual attraction between 
electrons moving in the same direction. When an electron is combined with a 
proton to produce a neutron and a neutrino, the required energy is in part 
the attractive energy due to all of the other parallel moving electrons 
in the wire albeit only one electron is actually destroyed. 
For an electron moving at a velocity \begin{math} v \end{math} uniform in time, 
the magnetic energy interaction of that single electron with the 
current \begin{math} I \end{math} due to all of the other electrons follows from 
Eq.(\ref{power}),
\begin{equation}
W_{magnetic}=-\eta mc^2\left(\frac{I}{I_0}\right)\frac{v}{c}\ .
\label{magnetic}
\end{equation}
The velocity of the electron is opposite to the direction of the current since 
electrons are negatively charged.

Let us now return to the energy considerations concerning the 
Wendt-Irion\cite{WendtIrion:1922} experiment.  
Typically, a capacitor discharge sent \begin{math} N\sim 2\times 10^{16} \end{math}  
electrons from one capacitor plate to the other capacitor plate. The initial 
energy in the capacitor was 
\begin{math} W_{Coulomb}\sim 15 \end{math} KeV per stored electron.
The total energy balance then dictates that at most 
\begin{math} N^* \sim 10^{13} \end{math} of these electrons could make inverse 
beta transitions causing nuclear transmutations. 
Many electrons acting cooperatively contribute energy 
\begin{math} W_{magnetic} \end{math}  
to inverse beta transitions even though only one of those electrons is destroyed. 
The Wendt-Irion peak current ratio \begin{math} I/I_0 \end{math}  was as high 
as two hundred\cite{{Wendt:1922}} yielding 
\begin{math} W_{magnetic}\sim 200\ {\rm MeV}\times v/c  \end{math}.
If the electron velocity in the filament is small, say 
\begin{math} v/c \sim 0.1  \end{math}, then \begin{math} W_{magnetic} \end{math}    
is more than sufficient for an inverse beta transition.
From the time duration of the pulse, one finds a mean drift velocity 
\begin{math} \bar{v}/c \sim 10^{-4}  \end{math}  of the electrons from one 
capacitor plate to the other capacitor plate through the transformer coil 
and the filament much smaller than the required electron velocity in the filament. 
To confirm the inequality \begin{math} \bar{v}\ll v \end{math}, 
consider an incompressible fluid moving between a thick pipe and a thin pipe. 
The fluid velocity is much larger in the thin pipe than the fluid velocity is in 
the thick pipe. A similar effect takes place for electrons flowing between 
thick circuit wires and thin wire filaments. Our energy considerations are now 
completed. 

Finally, for an electron beam in the vacuum there is a mutual electric Coulomb 
repulsion energy which overcomes a mutual magnetic Ampere attraction energy. The 
repulsion hinders the above cooperative many electron energy transfer for an 
inverse beta transition employing a vacuum beam as did Rutherford. 
It is only the Wendt wire filament, wherein the Coulomb repulsion between electrons 
is screened by positive ions, that enables nuclear transmutations.


\begin{thebibliography}{07}

\bibitem{WendtIrion:1922}
G.L. Wendt and C.E. Irion, {\it Amer. Chem. Soc.} {\bf 44}, (1922).

\bibitem{Rutherford:1922}
E. Rutherford, {\it Nature} {\bf 109}, 418 (1922).

\bibitem{Wendt:1922}
G.L. Wendt, {\it Science} {\bf 55}, 567 (1922).

\bibitem{Stephanakis:1972}
S. Stephanakis, et al., {\it Phys. Rev. Let.} {\bf 29}, 568 (1972).

\bibitem{Young:1977}
F. Young, et al., {\it J. Appl. Phys.} {\bf 48}, 3642 (1977).

\bibitem{Bakshaev:2001}
Y. Bakshaev et al., {\it Plasma Phys. Rep.} {\bf 27}, 1039 (2001).

\bibitem{Bakshaev:2006}
Y. Bakshaev et al., {\it Plasma Phys. Rep.} {\bf 32}, 501 (2006).

\bibitem{Velikovich:2007}
A. Velikovich et al., {\it Phys. Plasmas} {\bf 14}, 022701 (2007).

\bibitem{Coverdale:2007}
C. Coverdale et al., {\it Phys. Plasmas} {\bf 14}, 022706 (2007).

\bibitem{Sarkisov:2005}
G. Sarkisov, et al., {\it Phys. Plasmas} {\bf 12}, 052702 (2005).

\bibitem{Widom:2006}
A. Widom and L. Larsen, {\it Euro. Phys. J.} {\bf C46}, 107 (2006).













\end{thebibliography}
\end{document}